# PAIRWISE ALIGNMENT OF THE DNA SEQUENCE USING HYPERCOMPLEX NUMBER REPRESENTATION

**Jian-Jun Shu, and Li Shan Ouw**

School of Mechanical & Aerospace Engineering
Nanyang Technological University
50 Nanyang Avenue, Singapore 639798
e-mail: mjjshu@ntu.edu.sg, web page: http://www.ntu.edu.sg/home/mjjshu

**Abstract.** *A new set of DNA base-nucleic acid codes and their hypercomplex number representation have been introduced for taking the probability of each nucleotide into full account. A new scoring system has been proposed to suit the hypercomplex number representation of the DNA base-nucleic acid codes and incorporated with the method of dot matrix analysis and various algorithms of sequence alignment. The problem of DNA sequence alignment can be processed in a rather similar way as pairwise alignment of protein sequence.*

## 1 INTRODUCTION

Deoxyribonucleic acid, DNA, is the molecule of life. DNA is a double-helix comprising of two DNA strands running antiparallel to each other and is made of many units of nucleotides, which consist of a sugar, a phosphate, and a base. The four types of nucleotides (A, T, G and C) are linked in different order in the extremely long DNA molecules thus allowing unique DNA sequence for each of the infinite number of living organisms.

With more DNA sequences becoming available[1,2], computer programs have been developed to analyze these sequences in various ways. The dot matrix method, which is used to detect similarities between sequences, was discovered first[3]. In this method of comparing two sequences, a graph is drawn with one sequence written across a page from left to right and another sequence down the page on the left-hand side. A dot is placed where the corresponding nucleotide in the two sequences is the same. The graph is then scanned for diagonals of dots, which reveal similarities.

As the dot matrix method does not identify similarities that are interrupted, the method of sequence alignment was devised[4]. Sequence alignment is a procedure of comparing two sequences by searching for a series of individual characters that are in the same order. The best or optimal alignment is found after determining the type of sequence alignment desired. There are two types of sequence alignment: global alignment[5] and local alignment[6]. In the global alignment, the entire sequences are aligned from beginning to end. In the local alignment, parts of the sequences with the most matches are aligned, giving rise to a number of subalignments in the aligned sequences. These two methods of sequence comparisons are sometimes used hand-in-hand for more efficient sequence analysis of DNA. In this paper, the hypercomplex number system has been explored for its possible application in DNA sequencing.

## 2  HYPERCOMPLEX NUMBER REPRESENTATION

By permutation and combination, the total number of possible mixed DNA base-nucleic acid codes is $2^4 = 16$. Since there are four types of nucleotides, a four-dimensional space is essential to represent the DNA codes fully. The hypercomplex number system required here is a 3rd-order system of the form $\mathbf{z} = z_1 + z_2\mathbf{i} + z_3\mathbf{j} + z_4\mathbf{k} = (z_1, z_2, z_3, z_4)$. To assign the values for $z_{1,\cdots,4}$, the probability of each DNA base appeared in the DNA base-nucleic acid codes is taken into consideration. The values of $z_1$, $z_2$, $z_3$, and $z_4$ indicate the probabilities of the bases A, T, G, and C respectively, satisfying the basic principle that $z_1 + z_2 + z_3 + z_4 = 1$.

## 3  DNA BASE-NUCLEIC ACIDS IN HYPERCOMPLEX NUMBER REPRESENTATION

Based on the principle in the previous section, the hypercomplex number representation of the DNA base-nucleic acid codes are derived and listed in Table 1 below.

| Symbol | Meaning | Explanation | Hypercomplex number representation |
|--------|---------|-------------|------------------------------------|
| O | no base | no base | (0, 0, 0, 0) |
| A | A | Adenine | (1, 0, 0, 0) |
| T | T | Thymine | (0, 1, 0, 0) |
| G | G | Guanine | (0, 0, 1, 0) |
| C | C | Cytosine | (0, 0, 0, 1) |
| W | A or T | Weak interactions 2 h bonds | (½, ½, 0, 0) |
| R | A or G | puRine | (½, 0, ½, 0) |
| M | A or C | aMino | (½, 0, 0, ½) |
| K | G or T | Keto | (0, ½, ½, 0) |
| Y | C or T | pYrimidine | (0, ½, 0, ½) |
| S | C or G | Strong interactions 3 h bonds | (0, 0, ½, ½) |
| D | A, G or T not C | D follows C in alphabet | (⅓, ⅓, ⅓, 0) |
| H | A, C or T not G | H follows G in alphabet | (⅓, ⅓, 0, ⅓) |
| V | A, C or G not T | V follows U in alphabet | (⅓, 0, ⅓, ⅓) |
| B | C, G or T not A | B follows A in alphabet | (0, ⅓, ⅓, ⅓) |
| N | any base | any base | (¼, ¼, ¼, ¼) |

Table 1:    DNA base-nucleic acid codes and their hypercomplex number representation

## 4  DOT MATRIX WITH HYPERCOMPLEX NUMBER REPRESENTATION

In the dot matrix analysis using hypercomplex number representation of DNA bases, whether a dot is placed in a comparison of two DNA sequences is determined by dot product of the hypercomplex number representation of the DNA base-nucleic acids and the truncation value set. The probability of finding a match between the sequences is implied in dot product since the hypercomplex number representation assigned to each of the DNA base acids is based on the probability of each base appeared in Table 1. For instance, in a comparison of two sequences, an alignment between residues H and S, having the hypercomplex number representation of $(1/3, 1/3, 0, 1/3)$ and $(0, 0, 1/2, 1/2)$ respectively, the dot product value is derived as $\mathbf{z}_H \bullet \mathbf{z}_S = (1/3, 1/3, 0, 1/3) \bullet (0, 0, 1/2, 1/2) = 0.17$. In other words, based on dot product value (between $0$ and $1$) of the hypercomplex number representation of the residues in each sequence being compared, the truncation is set at the value of 1 (*i.e.*, any value less than 1 will be truncated to 0) for the conventional dot matrix analysis[3]. Unlike the conventional dot matrix[3], it is now a choice to set the truncation value for the desired stringency in finding a possible match: a higher value for higher stringency. For example, regions of short matching alignment may not be necessary. In order to prevent short diagonals from appearing too frequently and making the matrix too noisy to identify actual required aligned regions, a higher truncation value may be selected so as to reduce the number of dots between the two sequences. To illustrate the influence of various factors on the outcome of a dot matrix diagram, the following pair of sequences is selected as an example.

<div align="center">
T G R B W B H K M W C Y<br>
S Y A G M W D S H V R K
</div>

Using the above calculation, the dot product of the alignment between each residue of the example sequences is obtained and shown in a matrix in Figure 1.

|   | T | G | R | B | W | B | H | K | M | W | C | Y |
|---|---|---|---|---|---|---|---|---|---|---|---|---|
| S | 0 | 0.50 | 0.25 | 0.33 | 0 | 0.33 | 0.17 | 0.25 | 0.25 | 0 | 0.5 | 0.25 |
| Y | 0.5 | 0 | 0 | 0.33 | 0.25 | 0.33 | 0.33 | 0.25 | 0.25 | 0.25 | 0.5 | 0.5 |
| A | 0 | 0 | 0.5 | 0 | 0.5 | 0 | 0.33 | 0 | 0.5 | 0.5 | 0 | 0 |
| G | 0 | 1 | 0.5 | 0.33 | 0 | 0.33 | 0 | 0.5 | 0 | 0 | 0 | 0 |
| M | 0 | 0 | 0.25 | 0.17 | 0.25 | 0.17 | 0.33 | 0 | 0.5 | 0.25 | 0.5 | 0.25 |
| W | 0.5 | 0 | 0.25 | 0.17 | 0.5 | 0.25 | 0.33 | 0.25 | 0.25 | 0.5 | 0 | 0.25 |
| D | 0.33 | 0.33 | 0.33 | 0.22 | 0.33 | 0.22 | 0.22 | 0.33 | 0.17 | 0.33 | 0 | 0.17 |
| S | 0 | 0.5 | 0.25 | 0.33 | 0 | 0.33 | 0.17 | 0.25 | 0.25 | 0 | 0.5 | 0.25 |
| H | 0.33 | 0 | 0.17 | 0.22 | 0.33 | 0.22 | 0.33 | 0.17 | 0.33 | 0.33 | 0.33 | 0.33 |
| V | 0 | 0.33 | 0.33 | 0.22 | 0.17 | 0.22 | 0.33 | 0.17 | 0.33 | 0.22 | 0.33 | 0.17 |
| R | 0 | 0.5 | 0.5 | 0.17 | 0.25 | 0.17 | 0.17 | 0 | 0.25 | 0.25 | 0 | 0 |
| K | 0.5 | 0.5 | 0.17 | 0.33 | 0.25 | 0.33 | 0.17 | 0.5 | 0 | 0.25 | 0 | 0.25 |

Figure 1: The dot product values of hypercomplex number representation per aligned residue pair of the example sequences

## 5  EFFECT OF TRUNCATION VALUE ON DOT MATRIX ANALYSIS

Based on the dot product value per aligned residue of the example sequences, a comparison between the dot matrix diagram is shown in Figure 2, where the truncation values are at 0.3 and 0.5 respectively. The sequences are compared on a one to one residue basis. The dots are placed where the dot product values of the corresponding residues meet the designed truncation value.

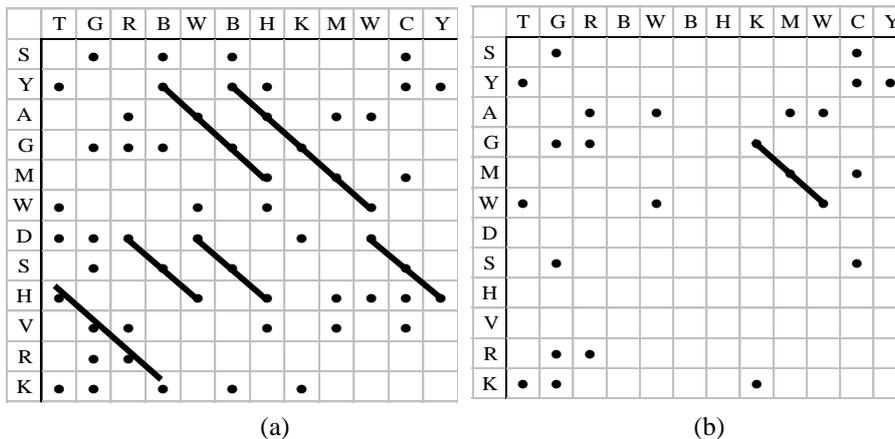

Figure 2: Dot matrix of example sequences with truncation values of (a) 0.3, and (b) 0.5

Varying the truncation value changes the number of dots appeared on the dot matrix diagram. For the case of Figure 2(a), many dots are presented, as the truncation value is set relatively low. The high concentration of dots on the diagram makes it deceive that there are many matched regions. However, after inserting diagonals, it is obvious that many dots are not collinear. They are only random matches all over the matrix. In addition, the number of aligned regions is also higher in (a) than in (b) as the truncation value indicates stringency in finding matches. With a lower truncation value in (a), we are actually looking for a higher number of possible matches even with a small probability than a more certain alignment as in (b), which has a higher truncation value.

## 6  SCORING MODEL FOR HYPERCOMPLEX NUMBER REPRESENTATION

A new scoring system is introduced by initially taking dot product of the DNA base hypercomplex number representation shown in Table 1, $\mathbf{x}_i \bullet \mathbf{y}_j = (x_{i_1}, x_{i_2}, x_{i_3}, x_{i_4}) \bullet (y_{j_1}, y_{j_2}, y_{j_3}, y_{j_4}) = x_{i_1} y_{j_1} + x_{i_2} y_{j_2} + x_{i_3} y_{j_3} + x_{i_4} y_{j_4}$. The new score values are then calculated using $s(\mathbf{x}_i, \mathbf{y}_j) = \mathbf{x}_i \bullet \mathbf{y}_j \times 20 - 5$, where the highest aligned score is $15$ and the lowest one is $-5$, with a gap penalty of $8$ for computational efficiency. After scaling the dot product value and rounding off to the nearest integer, the new scoring matrix is shown in Figure 3.

|   | A | T | G | C | W | R | M | K | Y | S | D | H | V | B | N |
|---|---|---|---|---|---|---|---|---|---|---|---|---|---|---|---|
|   | -5 | -5 | -5 | -5 | -5 | -5 | -5 | -5 | -5 | -5 | -5 | -5 | -5 | -5 | -5 |
| A | -5 | 15 | -5 | -5 | -5 | 5 | 5 | 5 | -5 | -5 | -5 | 2 | 2 | 2 | -5 | 0 |
| T | -5 | -5 | 15 | -5 | -5 | 5 | -5 | -5 | 5 | 5 | -5 | 2 | 2 | -5 | 2 | 0 |
| G | -5 | -5 | -5 | 15 | -5 | -5 | 5 | -5 | 5 | -5 | 5 | 2 | -5 | 2 | 2 | 0 |
| C | -5 | -5 | -5 | -5 | 15 | -5 | -5 | 5 | -5 | 5 | 5 | -5 | 2 | 2 | 2 | 0 |
| W | -5 | 5 | 5 | -5 | -5 | 5 | 0 | 0 | 0 | 0 | -5 | 2 | 2 | -2 | -2 | 0 |
| R | -5 | 5 | -5 | 5 | -5 | 0 | 5 | 0 | 0 | -5 | 0 | 2 | -2 | 2 | -2 | 0 |
| M | -5 | 5 | -5 | -5 | 5 | 0 | 0 | 5 | -5 | 0 | 0 | -2 | 2 | 2 | -2 | 0 |
| K | -5 | -5 | 5 | 5 | -5 | 0 | 0 | -5 | 5 | 0 | 0 | 2 | -2 | -2 | 2 | 0 |
| Y | -5 | -5 | 5 | -5 | 5 | 0 | -5 | 0 | 0 | 5 | 0 | -2 | 2 | -2 | 2 | 0 |
| S | -5 | -5 | -5 | 5 | 5 | -5 | 0 | 0 | 0 | 0 | 5 | -2 | -2 | 2 | 2 | 0 |
| D | -5 | 2 | 2 | 2 | -5 | 2 | 2 | -2 | 2 | -2 | -2 | 2 | -1 | -1 | -1 | 0 |
| H | -5 | 2 | 2 | -5 | 2 | 2 | -2 | 2 | -2 | 2 | -2 | -1 | 2 | -1 | -1 | 0 |
| V | -5 | 2 | -5 | 2 | 2 | -2 | 2 | 2 | -2 | -2 | 2 | -1 | -1 | 2 | -1 | 0 |
| B | -5 | -5 | 2 | 2 | 2 | -2 | -2 | -2 | 2 | 2 | 2 | -1 | -1 | -1 | 2 | 0 |
| N | -5 | 0 | 0 | 0 | 0 | 0 | 0 | 0 | 0 | 0 | 0 | 0 | 0 | 0 | 0 | 0 |

Figure 3: New scoring matrix derived from dot product of hypercomplex number representation of DNA bases

The conventional alignment algorithms[4] are used together with the hypercomplex number representation of the base pairs and the new scoring model introduced here. A pair of DNA sequences is used throughout the rest of this paper as a demonstration of the feasibility of this new scoring model:

H T A G A W M H R Y
T A W H C A M B H R

## 7  GLOBAL ALIGNMENT USING HYPERCOMPLEX NUMBER REPRESENTATION

Using the new scoring matrix in Figure 3, the following global dynamic programming matrix is set up using the example DNA sequence pair.

|   |   | H | T | A | G | A | W | M | H | R | Y |
|---|---|---|---|---|---|---|---|---|---|---|---|
|   | 0 | -8 | -16 | -24 | -32 | -40 | -48 | -56 | -64 | -72 | -80 |
| T | -8 | 2 | 7 | -1 | -9 | -17 | -25 | -33 | -41 | -49 | -57 |
| A | -16 | -6 | -1 | 22 | 14 | 6 | -2 | -10 | -18 | -26 | -34 |
| W | -24 | -14 | -1 | 14 | 17 | 19 | 11 | 3 | -5 | -13 | -21 |
| H | -32 | -22 | -9 | 6 | 9 | 19 | 21 | 13 | 5 | -3 | -11 |
| C | -40 | -30 | -17 | -2 | 1 | 11 | 14 | 26 | 18 | 10 | 2 |
| A | -48 | -38 | -25 | -2 | -7 | 16 | 16 | 19 | 28 | 22 | 14 |
| M | -56 | -46 | -33 | -10 | -7 | 8 | 16 | 21 | 21 | 28 | 22 |
| B | -64 | -54 | -41 | -18 | -8 | 0 | 8 | 14 | 20 | 20 | 30 |
| H | -72 | -62 | -49 | -26 | -16 | -6 | 2 | 10 | 16 | 18 | 22 |
| R | -80 | -70 | -57 | -34 | -24 | -11 | -6 | 2 | 8 | 21 | 14 |

Figure 4: Global dynamic programming matrix for hypercomplex number representation of DNA sequences

From the above matrix, the corresponding optimal alignment of the two sequences with a total score of $14$ is obtained as follows.

$$\begin{array}{c} \text{H T A G A W M H R Y} - - \\ - \text{T A} - \text{W H C A M B H R} \end{array}.$$

## 8  LOCAL ALIGNMENT USING HYPERCOMPLEX NUMBER REPRESENTATION

Using the same pair of DNA sequences with hypercomplex number representation, the local dynamic programming algorithm is implemented to give the following matrix.

|   |   | H | T | A | G | A | W | M | H | R | Y |
|---|---|---|---|---|---|---|---|---|---|---|---|
|   | 0 | 0 | 0 | 0 | 0 | 0 | 0 | 0 | 0 | 0 | 0 |
| T | 0 | 2 | 15←7 | 0 | 0 | 5 | 0 | 2 | 0 | 5 |
| A | 0 | 2 | 7 | 30←22 | 15←7 | 10←2 | 7 | 0 |
| W | 0 | 2 | 7 | 22 | 25 | 27 | 20←12 | 12←14 | 7 |
| H | 0 | 2 | 4 | 14 | 17 | 27 | 29←21 | 14 | 10 | 16 |
| C | 0 | 2 | 0 | 6 | 9 | 19 | 22 | 34←26←18 | 15 |
| A | 0 | 2 | 0 | 15←7 | 24 | 24 | 27 | 36 | 31←23 |
| M | 0 | 2 | 0 | 7 | 10 | 16 | 24 | 29 | 29 | 36 | 31 |
| B | 0 | 0 | 4 | 0 | 9 | 8 | 16 | 22 | 28 | 28 | 38 |
| H | 0 | 2 | 2 | 6 | 1 | 11 | 10 | 18 | 24 | 26 | 30 |
| R | 0 | 0 | 0 | 7 | 11 | 6 | 11 | 10 | 16 | 29 | 22 |

Figure 5:  Local dynamic programming matrix for hypercomplex number representation of DNA sequences

In the local dynamic programming matrix, it is not necessary to start the alignment at the bottom right cell. Instead, the alignment starts at the cell with the highest score so that the optimal local alignment can be found. In this case, the highest score is $38$. Thus the traceback starts from there and ends when it reaches a score of $0$. The optimal local alignment of this pair of example sequences has a score of $38$ and is found to be

$$\begin{array}{c} \text{T A G A W M H R Y} \\ \text{T A} - \text{W H C A M B} \end{array}.$$

## 9  REPEATED MATCHES USING HYPERCOMPLEX NUMBER REPRESENTATION

Likewise, by applying the algorithm for repeated matches with the new scoring model to the example sequences, the same DNA sequences demonstrates the outcome. Two different values of T score are used to show the influence of the threshold value on the resultant optimal alignment.

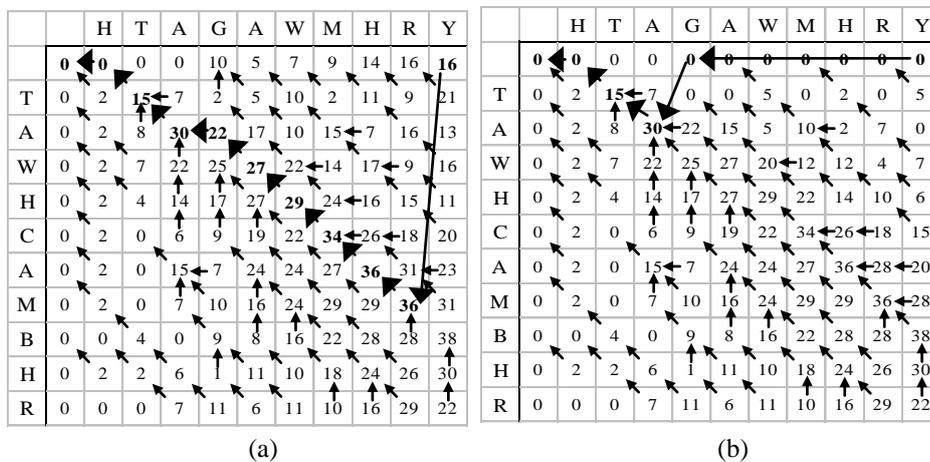

(a)            (b)

Figure 6:  Repeat dynamic programming matrix for hypercomplex number representation of DNA sequences with threshold scores of (a)  $T = 20$  and (b)  $T = 40$

For a threshold value of $20$, the optimal alignment is

$$\begin{array}{cccccccccc} H & T & A & G & A & W & M & H & R & Y \\ - & T & A & - & W & H & C & A & M & \end{array}.$$

For a threshold value of 40, a different optimal alignment is obtained as

$$\begin{array}{cccccccccc} H & T & A & G & A & W & M & H & R & Y \\ - & T & A & . & - & - & - & - & - & - \end{array}.$$

When the threshold value is increased significantly, a large portion of the sequence is excluded from the matched region. In other words, a larger threshold score implies a higher stringency.

## 10 OVERLAP MATCHES USING HYPERCOMPLEX NUMBER REPRESENTATION

To find out whether the example hypercomplex DNA sequences show traces of overlapping, they are subjected to the same overlapping dynamic programming. A threshold of 20 is pre-specified.

|   |   | H | T | A | G | A | W | M | H | R | Y |
|---|---|---|---|---|---|---|---|---|---|---|---|
|   | 0 | 0 | 0 | 0 | 0 | 0 | 0 | 0 | 0 | 0 | 0 |
| T | 0 | 2 | 15 | 7 | -1 | -5 | 5 | -3 | 2 | -5 | 5 |
| A | 0 | 2 | 7 | 30 | 22 | 14 | 6 | 10 | 2 | 7 | -1 |
| W | 0 | 2 | 7 | 22 | 25 | 27 | 19 | 11 | 12 | 4 | 7 |
| H | 0 | 2 | 4 | 14 | 17 | 27 | 29 | 21 | 13 | 10 | 6 |
| C | 0 | 2 | -3 | 6 | 9 | 19 | 22 | 34 | 26 | 18 | 15 |
| A | 0 | 2 | -3 | 12 | 4 | 24 | 24 | 27 | 36 | 31 | 23 |
| M | 0 | 2 | -3 | 4 | 7 | 16 | 24 | 29 | 29 | 36 | 31 |
| B | 0 | -1 | 4 | -4 | 6 | 8 | 16 | 22 | 28 | 28 | 38 |
| H | 0 | 2 | 1 | 6 | -2 | 8 | 10 | 18 | 24 | 26 | 30 |
| R | 0 | -2 | -3 | 6 | 11 | 3 | 8 | 10 | 16 | 29 | 22 |

Figure 7: Overlap dynamic programming matrix for hypercomplex number representation DNA sequences with threshold of 20

The possible overlap matching sequence is shown below. The optimal overlapping alignment has a score of 38. The resulting alignment is the same as that obtained for local alignment in the earlier section but this is not always true for other sequences.

$$\begin{array}{cccccccc} T & A & G & A & W & M & H & R & Y \\ T & A & - & W & H & C & A & M & B \end{array}.$$

## 11 CONCLUDING REMARKS

To represent fully the DNA base-nucleic acid codes in hypercomplex number, a four-dimensional space is required. The representation number assigned to each base code takes into consideration the probability of each nucleotide in the DNA code. The conditions assumed in the assignment of the representation are that the probabilities for the occurrences of A, T, G and C are equal and the sum of the individual probabilities is 1.

The implementation of hypercomplex numbers in the dot matrix method brings forth an improvement to the conventional method[3] of placing a dot when there is a match between the corresponding residues of two sequences. As the hypercomplex number representation of DNA base-nucleic acid codes is in numbers instead of alphabetical characters, the significance of probabilistic sequencing is emphasized. To determine whether a dot should be placed between the aligned residues, dot product of the hypercomplex number representation of the bases is taken and truncated. With the introduction of 'value' instead of 'dots' as in the conventional method[3], the truncation value can be varied and hence a greater control over the degree of alignment desired is possible. A higher truncation value corresponds to a higher stringency for longer matching regions between the sequences.

To use the hypercomplex number representation of DNA sequences, a new scoring model has been derived. The new model, with the consideration of probability of each nucleotide presented in the DNA base-nucleic acid codes, uses the dot product arithmetic between the residues of the sequences to be matched. The dot product value is scaled and rounded off to an integer. The various algorithms have been applied to the sample sequence in hypercomplex number representation and the feasibility of using the hypercomplex number representation and scoring model has been verified. As most of the DNA codes consisting of mixed bases, the alignments obtained for the various algorithms are very high. This is because the algorithms can detect a possible alignment with small possibility of a match between the two sequences.